\documentclass{iau307}
\usepackage{graphicx}
\usepackage{natbib}
\usepackage{url}
\usepackage{dtklogos}
\bibpunct{(}{)}{;}{a}{}{,}

\title[Hydrodyanmic Models of Massive Stars] 
{Behaviour of Pulsations in Hydrodynamic Models of Massive Stars}

\author[Lovekin \& Guzik]   
{C.C. Lovekin$^1$
 \and J.A. Guzik$^2$}

\affiliation{$^1$Department of Physics, Mount Allison University, Sackville, NB, Canada\\ email: {\tt clovekin@mta.ca} \\[\affilskip]
$^2$XTD-NTA, Los Alamos National Laboratory, Los Alamos, NM 87545}

\pubyear{2014}
\volume{307} 
\pagerange{}
\jname{New windows on massive stars: asteroseismology, interferometry, and spectropolarimetry}
\editors{G. Meynet, C. Georgy, J.H. Groh \& Ph. Stee, eds.}

\begin{document}

\maketitle

\begin{abstract}
We have calculated the pulsations of massive stars using a nonlinear hydrodynamic code including time-dependent convection.  The basic structure models are based on a standard grid published by Meynet et al. (1994).  Using the basic structure, we calculated envelope models, which include the outer few percent of the star.  These models go down to depths of at least 2 million K.  These models, which range from 40 to 85 solar masses, show a range of pulsation behaviours.  We find models with very long period pulsations ( $>$ 100 d), resulting in high amplitude changes in the surface properties.  We also find a few models that show outburst-like behaviour. The details of this behaviour are discussed, including calculations of the resulting wind mass-loss rates.
\end{abstract}

\firstsection 
\section{Introduction}

S Doradus variables, or Luminous Blue Variables (LBVs) show photometric variability on several different timescales. The shortest, the ``microvariability'' has timescales of weeks to months.   For example, \citet{vanG98} show the photometric variability of R85, which is a B5Iae $\alpha$ Cyg variable and quite possibly also a LBV.   The light curve varies by about 0.2 to 0.3 magnitudes on timescales of about 100 days.  Their best fit solution found two periods, one at 390 d and one at 83.5 d \citep{vanG98}.  

These stars periodically undergo larger outbursts, during which the visual magnitude of the stars increases while the bolometric magnitude stays constant.  This process recurs on either short ($<$ 10 years) or long ($>$ 20 years) timescales.  The mechanism that produces these events is unclear, although it is thought to be related to their proximity to the Humphreys-Davidson (HD) limit \citep{hd94}.  As these hot massive stars evolve towards the red side of the HR diagram, they encounter the HD limit and become unstable, then undergo a period of mass loss before settling back on the stable (hot) side of the HD limit.  This process repeats for a period of a few times $10^4$ years.  
 
 LBVs can also undergo giant eruptions as observed in $\eta$ Carinae.  These events can eject large quantities of mass (10 M$_{\odot}$ or more during the Great Eruption of $\eta$ Car), and produce an increase in both bolometric and visual luminosity \citep{smith}.  

In this work, we investigate the possibility that the interaction between radial pulsations and time-dependent convection can drive the long-period variability observed in these stars.  A few of our models undergo outburst-like events that may be related to the outbursts seen in S Doradus variables.  We also present preliminary estimates of the mass-loss rates in these stars.

\section{Models\label{models}}

Our models are based on the stellar evolution model grid calculated by \citet{mm94}.  These models include enhanced mass-loss rates on the post-main sequence, but do not include rotation.  In this work, we focus on the 60 and 85 M$_{\odot}$ models at metallicities of 0.004, 0.008, 0.02, and 0.04. We selected models along each track ranging from approximately the middle of the main sequence through the start of core helium burning and used the current mass, effective temperature, luminosity, and composition at the surface and in the core to calculate an envelope model.  Our envelope models contain 60 zones, including the outer few percent of the mass, and go down to a depth of $2\times10^6$ K, which ensures that we have completely captured the damping and driving regions.  For more details on the model calculations, refer to \citet{LGa,LGb}.  

We use the DYNSTAR hydrodynamic code, which includes time-dependent convection (TDC) \citep{ostlie}.   This is a more realistic approximation than the standard mixing length model when the timescales of the pulsations are similar to the convective timescale, as is the case here.  Because the convective motions increase gradually after a region becomes convectively unstable, energy can be trapped in the lower layers, causing the radiative luminosity in these regions to surpass the Eddington limit.  In models with convective regions, we expect to see longer periods in the hydrodynamic model, as well as other behaviour not predicted by the linear non-adiabatic pulsation calculations.  Our model of TDC is described more fully in \citet{LGa}, and builds on previous work \citep[see][]{GL}.  

We calculate the pulsation periods using a linear non-adiabatic pulsation code and DYNSTAR.  The pulsation periods are extracted from the nonlinear results as described in \citet{LGa}.  A comparison of the resulting periods shows that the nonlinear periods can be longer than the linear periods by factors of 50 or more, while few models have nonlinear periods that are longer by a factor of 1000 or more. Models with convection zones show larger period enhancements than models without convection, suggesting that the interaction between pulsation and TDC produces longer periods than prediced by the linear model alone.  

\section{Long Period Pulsations\label{longp}}

\begin{figure}[b]
\begin{center}
\includegraphics[width=\textwidth]{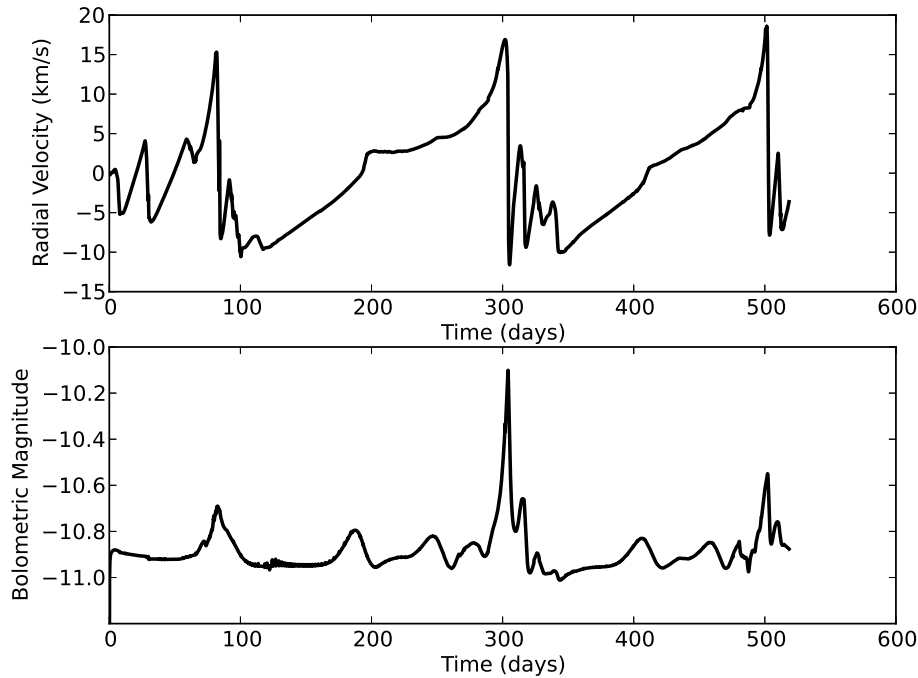} 
\caption{Top: The radial velocity variation for a typical model pulsating with a long period.  This model is an 85 M$_{\odot}$ model with metallicity Z = 0.004 near the end of the main sequence. Bottom:  The bolometric magnitude variation.  Variations are typically about 0.2 magnitudes, as seen in R85 and other S Dor variables.}
\label{fig1}
\end{center}
\end{figure}

Plotting the location of models with periods $>$ 50 days shows an instability strip that is consistent with the location of the HD limit in the HR diagram.  The models in this region are typically near the end of their main-sequence lifetime and are mainly located to the red of the HD limit. The resulting variability is typically multiperiodic, with a dominant period on the order of 100 days.  One of the longest lightcurves calculated is over 15\,000 days, although simulation times of a few hundred days, capturing 2-3 pulsation cycles are more typical, as shown in the Figure \ref{fig1}.  The peak-to-peak amplitude of this pulsation is on the order of 30 km/s.  Given the long pulsation period, the expansion and contraction produces huge changes in effective temperature ($\sim$ 9\,000 K) and radius ($\sim$ 600 R$_{\odot}$).  The resulting magnitude changes, shown in the bottom half of Figure \ref{fig1}, are typically on the order of 0.2 magnitudes, consistent with those seen in R85 \citep{vanG98}.

\section{Outbursts\label{outbursts}}

\begin{figure}[b]
\begin{center}
\includegraphics[width=\textwidth]{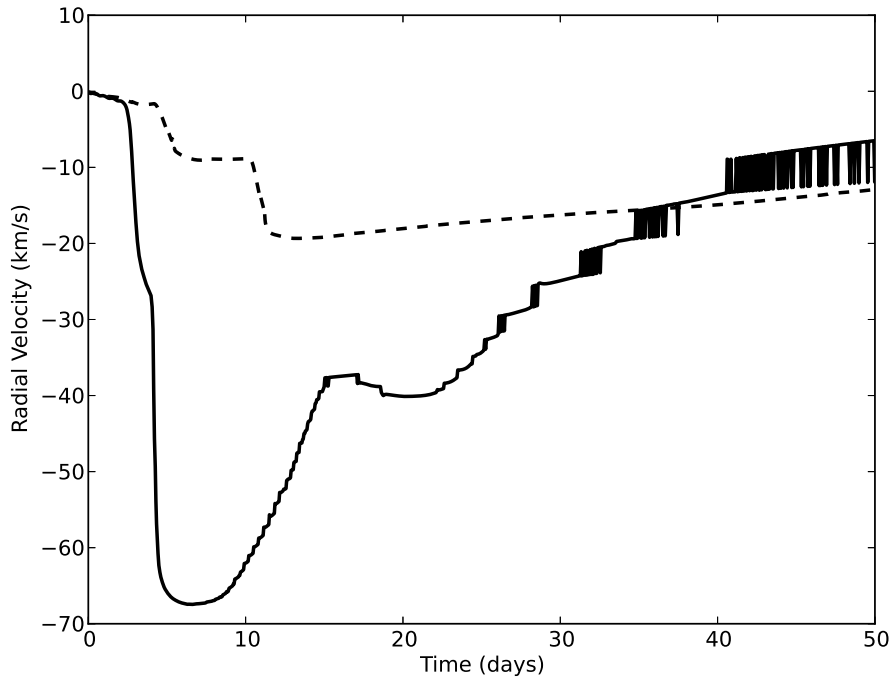} 
\caption{The radial velocity curves for a major (solid) and minor (dashed) outburst.  Minor outbursts occur later in the simulation than major outbursts, and have significantly lower peak expansion velocities.  The calculation after $\sim$ 15-20 days is uncertain, as our simulations do not remove mass from the envelope.  Although the expansion velocities are significantly lower than the escape speed for these stars, the outbursts are expected to drive an increase in mass loss (see Section \ref{massloss}).}
\label{fig:outbursts}
\end{center}
\end{figure}

A few of our models show outburst-like events during which the surface expands rapidly in the first few days of the simulation.  We divide these events into two classes, based on the peak expansion speed of the surface.    Minor outbursts have peak expansion velocities of 20-30 km/s, while the major outbursts have peaks between 50 and 80 km/s.  The radial velocity curve of a typical example of each class is shown in Figure \ref{fig:outbursts}. Even in the major outbursts, the expansion velocities are not high enough to eject mass, but the resulting variation in surface parameters will likely drive increased mass loss.  Unfortunately, our hydrodynamic calculations do not allow us to remove mass from the star during the calculation, so we cannot reliably follow the behaviour of the star after the outburst.

In our models, outbursts occur preferentially in models with lower metallicity.  The model showing the major outburst illustrated in Figure \ref{fig:outbursts} has Z = 0.004, and the other outbursting models are all either Z = 0.004 or Z = 0.008.  This is a result of the high mass-loss rates included in these models, as higher metallicity models do not reach the region of the HR diagram where outbursts are common.

\subsection{Minor Outbursts}

Minor outbursts have peak expansion velocities of 20-30 km/s, which is comparable to the long-period pulsation amplitudes.  Minor outbursts typically occur after approximately 10 days of simulation time.  As the envelope expands and contracts during pulsation, the interior becomes convectively unstable.  Because of the TDC, the convective velocities, and hence the convective luminosities increase gradually, with the increase in luminosity lagging behind the increase in velocity by about 0.5-1 day.  This pattern is first seen in the deepest zone we followed (about 1/2 way into the envelope), and then gradually propagates out to the surface.  When the peak in the convective luminosity reaches the surface, this drives a sudden jump in the expansion rate of the surface.  As a result, the effective temperature and radius increase by 7\,000 K and 200 R$_{\odot}$ respectively.  During this time the bolometric magnitude remains approximately constant.  

\subsection{Major Outbursts}

Major outbursts occur earlier in the simulation than the minor outbursts, typically around 5 days.  In these models, the envelope is initially convective, but then the convection turns off as the star pulsates.  When the convection turns off, the radiative luminosity spikes, jumping up to 20-25 times the Eddington luminosity.  This sharp increase in the radiative luminosity drives the expansion of the star, reaching peak expansion  velocities of up to 80 km/s.  This pattern results in major changes in the temperature and radius.  During the spike in luminosity, the temperature jumps by more than 10\,000 K, then drops by nearly 20\,000 K as the star expands.  The spike in luminosity and temperature also corresponds to a brightening of 3 in bolometric magnitude.

\section{Mass-Loss Rates\label{massloss}}

We have used the surface properties as a function of time to calculate mass-loss rates using the \citet{vink01} mass-loss prescription.  These results are highly uncertain, as the Vink rates are based on calculations of line-driven winds in main sequence stars, while our models include post-main sequence stars  and the winds are likely to be continuum-driven.  Given the rapid variation in the surface parameters of our models however, it is not clear that any simple prescription is valid, so we have chosen the Vink rates for ease of use.  It is our expectation that while the absolute values of the resulting mass-loss rates are inaccurate, the relative changes in rate are reasonable.  

For the long period model shown in Figure \ref{fig1}, the mass-loss rate varies between 10$^{-8}$ and 10$^{-12}$ M$_{\odot}$/yr, a variation of 4 orders of magnitude.  The mass-loss rates calculated for this model are considerably lower than expected for an 85 M$_{\odot}$ star, which is an indication that the mass-loss rates we use here are not a good choice.  This particular model has (log L/L${\odot}$, log T$_{eff}$) = (6.151, 4.498), which places it just past the end of the main sequence, where the Vink mass-loss rates are not expected to be accurate.

For the major outburst shown in Figure \ref{fig:outbursts}, we find that the mass-loss rate increases dramatically during the sharp spike in radiative luminosity.  The mass-loss rate increases by nearly 5 orders of magnitude, from $\sim$ 10$^{-8}$ M$_{\odot}$/yr to $\sim$ 10$^{-3}$ M$_{\odot}$/yr before dropping as the star expands.  The peak mass-loss rate in this model is in very good agreement with the observed mass-loss rates during S Doradus type outbursts.  However, this model is slightly cooler than the long-period model discussed above, with (log L/L${\odot}$, log T$_{eff}$) = (6.195, 4.168), putting this star even further past the main sequence.  As a result, the Vink mass-loss rates are not expected to be appropriate for this star, and we should only consider the changes in mass-loss rate to be reasonable.  

\section{Conclusions\label{conclusions}}

We have performed hydrodynamic calculations of massive pulsating stars including the effects of time-dependent convection (TDC).  We have found that TDC  interacts with the pulsation to significantly alter the pulsation characteristics, producing periods that are tens or even thousands of times longer than predicted by linear non-adiabatic calculations.  Even at relatively low amplitudes (10-15 km/s) this variability can act over periods of hundreds of days to produce very large changes in the radius, luminosity, and effective temperature of the stars.  The resulting magnitude changes are typically 0.2-3 magnitudes, similar to those observed in hot massive stars such as R85.  Simple mass loss calculations have shown that the mass-loss rate in these stars could vary by as much as 4 orders of magnitude over the pulsation cycle.  

A small subset of our models show outburst-like behaviour, classified based on the peak expansion velocity as either major (50-80 km/s) or minor (20-30 km/s). Even the highest peak speed in the major outbursts is well below the escape speed for these stars, but the mechanism that produces the outburst also produces a large increase in luminosity and temperature.  As a result, the star brightens by 3 magnitudes for about a day before the star begins to expand.  The peak mass-loss rate at this point is comparable to the mass-loss rates in S Dor variables, although the duration in our models is shorter.

\bibliographystyle{iau307}
\bibliography{IAUS307_Lovekin}

\begin{discussion}

\discuss{Lobel}{The microvariability of LBVs observed in V is typically a few tenths of a magnitude with quasi-periods from days to months. Do the hydro-models show that they can be attributed to pulsations or time-dependent convection, or both?}

\discuss{Lovekin}{The pulsations seen in our models are produced by an interaction between pulsations and time-dependent convection. Without the convection, the periods would be much shorter. The periods we find are typically larger than expected for the LBV microvariability, but that's somewhat artificial. For this work, we used an arbitrary cut-off of 20 days and only considered models with larger periods. It would be interesting to go back and look at the shorter period models in more detail.}

\discuss{de Koter}{When the LBVs inflate on a years-decades timescale they all end up with $T_\mathrm{eff}\sim 8000\,\mathrm{K}$, regardless their luminosity. This implies that the radius inflation factor is mass (or temperature) dependent. Do your predictions also show such a mass (or $T_\mathrm{eff}$) dependent inflation?}

\discuss{Lovekin}{This is something we haven't looked at. This could be very interesting and I'll definitely a look into this.}
\end{discussion}

\end{document}